\documentstyle[prl,aps]{revtex}
\input epsf.tex
\newcommand{\BE}{\begin{equation}}
\newcommand{\EE}{\end{equation}}
\newcommand{\BA}{\begin{eqnarray}}
\newcommand{\EA}{\end{eqnarray}}
\begin{document}
\draft

\title {Speed of light measurement using ping}
\author{Joel Lepak and M. Crescimanno}
\address{Center for Photon Induced Processes, 
Department of Physics and Astronomy, 
Youngstown State University, Youngstown, OH, 44555-2001}
\medskip

\date{\today} \maketitle

\begin{abstract}
We report on a very simple and inexpensive
method for determining the speed of 
an electrical wave in a transmission line. The method
consists of analyzing the roundtrip time for ethernet packets between 
two computers. It involves minimal construction, straightforward mathematics
and displays the usefulness of stochastic resonance in signal recovery.  
Using basic electrical properties of category-five 
cable students may use their measurements to determine
the speed of light in the vacuum to within a few percent. 

\vskip .2in
\end{abstract}
\noindent {\bf I. INTRODUCTION:}
With the now ubiquitous network and computer hardware in 
American secondary schools, high school and college
students have become quite familiar with the internet. 
But the fact that signals on the internet cables
typically move at 
hundreds of millions of  meters
per second (about 2/3 of the speed of light) 
is completely lost in the page download-and-rendering
time that students experience. This laboratory 
experience grew out of a desire to leverage the now ubiquitous
computer and network resources in schools for a simple, cheap, quality  
laboratory in which students measure
one of the most consequential natural 
constants, the fantastic speed of light.

The measurement of the speed of light$^{\cite{speeddefn}}$ has a history 
that students generally find interesting$^{\cite{asimov}}$. 
There are the early speculations about light 
of the philosophers followed by the failed attempts of the giants
of physics, Galileo and Newton. 
The first measurement of the speed of light was 
by Danish Astronomer Olaus Roemer in 1675 who 
used the moons of Jupiter as a clock. He determined the speed to be 
some 3/4 of its (now defined) speed, and a more accurate 
(about 5\%) measurement 
had to wait half a century until 
British Astronomer James Bradley in 1728 determined it
using the aberration of light from distant stars caused by the change
in the orbital velocity of the Earth over the course of the seasons. 
The first non-astronomical measurement of light's speed is due to 
French physicist Jean-Bernard-L\'eon Foucault in 1849 using the (now classic) 
quickly rotating mirror setup used later notably by 
American physicist Albert Michelson. 
Foucault's measurement was within a percent of the 
modern (defined) value. This time period 
was one of intense scientific scrutiny
of electricity and magnetism. In 1856 German physicists 
W. Weber and F. Kohlrausch used Leyden jars and a ballistic galvanometer 
to determine a ratio of electric and magnetic coupling strengths that 
has the dimensions of speed, and,  intriguingly, found that speed to be
near Foucault's value for light. All of this development preceded 
J. C. Maxwell's $^{\cite{maxwell}}$ 
theoretical synthesis of light as an electromagnetic wave
in 1865. In our physics 
laboratory the determination of the speed of light is a classic 
experiment involving quickly rotating mirrors 
or, more popular 
nowadays, a pulsed source$^{\cite{maryjames}}$ or a determination 
of the wavelength 
and frequency of microwaves. 

All of the many methods of determining the speed of light in common use 
in our introductory physics laboratory
and discussed in the literature 
(See for example Refs.~\cite{maryjames} \cite{laserpointer} \cite{tvghost}) 
have their experimental and 
pedagogic difficulties. In the experiment we chose and that we describe 
below our goals included

\medskip
\noindent (1) {\bf Simple Equipment:} The setup would require minimal 
effort and a minimal use of dedicated, specialized laboratory equipment. 

\medskip
\noindent (2) {\bf Simple Analysis:} The structure and meaning of the 
laboratory and the analysis of the data should be straightforward. 

\medskip
\noindent (3) {\bf Flexible:} The design should incorporate elements that
may be interchanged and manipulated to allow students to be creative. 

\medskip
\noindent (4) {\bf Time:} It should be possible to complete the 
data collection for a measurement in a single two hour period. 

\medskip
\noindent (5) {\bf Cost:} The apparatus should be of minimal cost. 

\medskip
\noindent (6) {\bf Quality:} The experiment must provide a determination 
of the speed of propagation (in air or in cable) with errors that are 
consistently not more than several percent.
\medskip

We chose a time-of-flight approach because it is straightforward both in 
meaning and in analysis. For example, no reference is made to 
waves or the wave nature
of light and thus this laboratory 
can logically be done early in the second semester sequence. 
The approach is simply to reflect
small data packets between two computers that are connected with 
ethernet card/cables and record the round trip time. 
This has the great advantage of being easy to set up and 
inexpensive, a few tens
of dollars for the the cost of a few
``category-5''  (hereafter referred to as ``cat-5'') 
and RG-58/U coaxial cables of different lengths. 
Such echo time signatures 
(in engineer's parlance 'Time Domain Reflectometry' or TDR) 
are used in sophisticated and expensive 
commercial network analysis equipment and this 
laboratory represents a very simplified version of these tools. 
It opens classroom discussion to a world of interesting 
questions regarding the physical fabric of the internet. 

The disadvantages of this method are (1) the additional 
abstraction of the theoretical connection between cable 
radio waves (invisible light) and free space waves and 
(2) the fact that the results are not immediately apparent from run to 
run but involve some (again straightforward) data analysis to see
the effect of changing the cable length. 

In addition to the advantages presented by our goals described above, 
other advantages of this method are
(1) it will further familiarize students with {\it ping} as a useful 
IP protocol and (2) it makes 
use of existing equipment (the computers and 
cable resources) that is easy to bring together for this laboratory yet 
of use elsewhere after the laboratory. This may be of 
particular advantage for high school physics class budgets. 

Finally,   
the experiment can be understood at several different 
levels of depth. At the most superficial level, this laboratory
excersizes
students' knowledge of the relation $d=vt$ and scientific notation. 
But on a deeper level, this laboratory introduces students to the 
idea of noise-assisted sub-threshold signal detection, which, in 
the modern physics jargon is called {\it stochastic resonance}. 

Here's the problem; 
the cable range of ethernet without a repeater is about 250 feet (that is, 
at most a few microseconds roundtrip in cat-5 cable) and actually all the 
tests described here are done in much shorter cat-5 cable 
(more practical for typical reuse) 
and coaxial cable lengths so that it 
can be done cheaply. A typical classroom can hold
several experiments of this type, the cables being shared 
between pairs of computers (and thus lab groups).  
Since {\it ping} only returns 
roundtrip times as measured in microseconds the actual signal (which 
is the additional delay in a cable path of longer length) is
below the (reported) resolution of {\it ping}. 

The solution is to use noise. 
Although noise usually hampers one's ability to measure a signal, 
in this experiment, noise in the form of randomly distributed 
small delays (microseconds) associated with machine response 
actually makes the measurement of the signal (nanosecond-long 
cable transit delays) possible. Without the noise, the experiment 
we describe here would be impossible! This 
concept of noise-assisted sub-threshold signal detection (hereafter; 
stochastic resonance) is of great value because it plays a role in 
a great variety of systems. 
For a readable introduction and overview of stochastic resonance
see Ref.~\cite{srreview} and Ref.~\cite{websitesr} for a bibliography. 
For example, stochastic resonance
has been used to analyze climate patterns$^{\cite{climate}}$ and 
plays a role in fundamental 
neuro-physiology$^{\cite{neuro}}$. 
Part of the hidden pedagogic agenda of this laboratory is 
to introduce the concept of stochastic resonance in a 
hands-on way. How well this laboratory can actually get students to ponder 
that depends on the approach of the instructor. Our experience with this
laboratory indicates that 
time differences on the order of 50 nanoseconds 
(or about 5 \% of the threshold) are reliably resolvable. 

\bigskip

\noindent {\bf II: Hardware, Software and Experimental Design} 

{\bf Hardware:} The hardware requirements of this experiment are basically two 
computers with ethernet NIC (Network Interface Cards) 
cards and various cable lengths. 
There is no need for any network switches or hubs. 
We have used laptops, some with built in 
ethernet NIC and others with PCMCIA NIC and both work fine. 
Although we have not tried this experiment on desktop machines we 
are confident that the data quality is largely independent of the 
hardware. 

However, besides portability, laptops are useful because we did 
learn that the return times depended somewhat on 
the line voltage. We ran the 
experiment only when the laptops were fully warmed up, 
charged and yet still plugged into 
the wall to ameliorate dependence on line/power variability. 

For cabling, we used standard cat-5 cables and RG-58/U coax of various length. 
Both of these types of cables are reasonably inexpensive and can
be used elsewhere in the laboratory and classroom. Finally, 
the lengths of cable used in this experiment are actually rather short.
The longest cable we used was 60m, though reasonably good results in 
cat-5 can be done with half that (see Figure 6).  
We have tried to repeat this experiment in cat-3 cable and 
were unable to get a reproducible signal presumably due to the 
noisy EMI environment (cat-3 is not as 
tightly twisted and subject to much more cross-talk and environmental 
electrical noise). For more technical information about cables and
ethernet relevant to this experiment consult Appendix A.

A crossover cable
allows one to create a direct connection between two machines, obviating 
the need (and expense) of a hub/switch for this experiment. Of course, 
such cables are also quite handy when moving 
files between two machines that are not on the same network. 

Alternatively, one can also use a hub or a switch and just straight-through 
cables of different lengths. We have tested that the experiment can be 
profitably done through a hub; our hub seemed to introduce an additional
25 microseconds of delay in the roundtrip time, but any timing 
noise associated with it's activity apparently was negligible at the 
timing threshold. 

Terminated cat-5 cables are typically sold in the 'straight-through'
configuration, that is, they are not cross-cabled, having instead
identically terminated ends. Thus, for the purposes of this 
experiment we made a single 
short (about 2.6 meter) cross-cabled cat-5 gender
changer (male on one end, female on the other) cable that we used in series
with commercial straight-through cat-5 cable of various 
lengths. 

For the second part of the experiment we measured the speed of 
propagation in a RG-58/U (standard 50 $\Omega$ coaxial cable) by 
again using different cable lengths. As described above, the reason 
for doing this was that the conversion from the measured speed to the 
speed of light in vacuum was easy to do by measuring geometrical 
and simple electrical characteristics of the cable. In terms of
hardware, we simply added BNC females at a break in a
single pair (orange or green) of a short (.95m) straight-through 
ethernet cable. This cable's RJ45 ends were then plugged into the 
crossover gender changer and the BNC females were mated with BNC males 
on the RG-58/U cables of various lengths. Additionally, we did find it 
necessary to impedance match the cat5 onto the coax; the 
coax has an impedance of $50 \Omega$ and cat-5 cable of 
$100 \Omega$. Figure 1 shows a schematic of the impedance matching 
circuit and a diagram of the shielded box used. Note that this arrangement 
allowed for only a rather short (about .05m) disturbance in the cat-5 
cable. 

\begin{figure}
\begin{center}
\epsfxsize=5in\epsffile{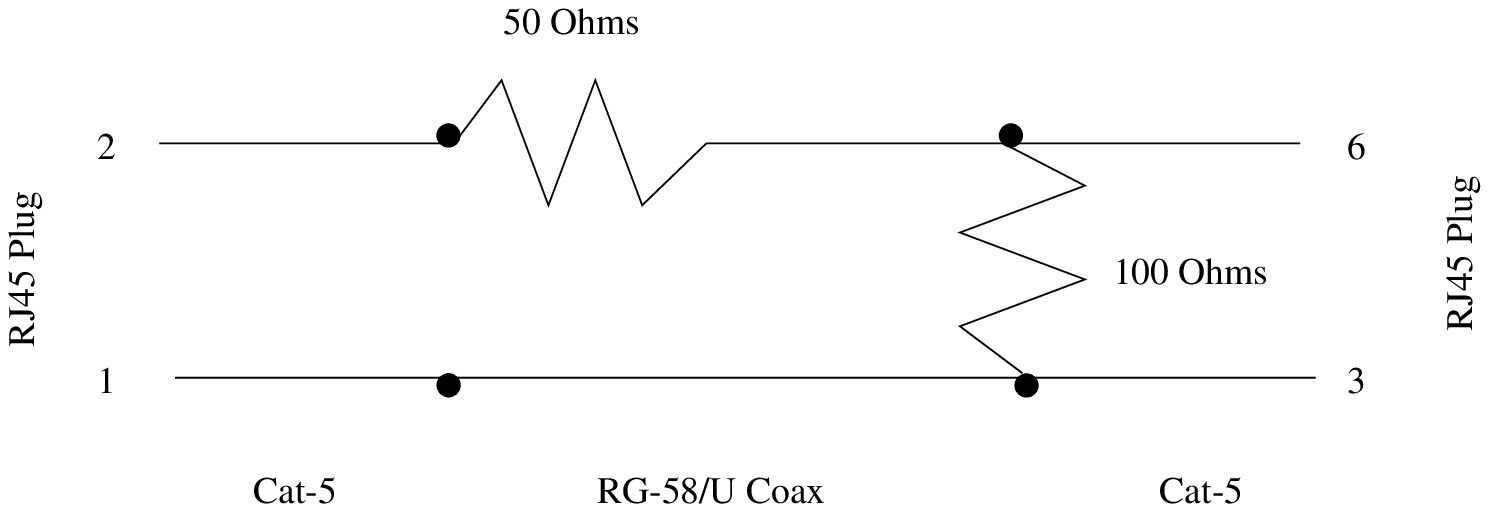} 
\label{fig1}
\end{center}
\end{figure}
\begin{figure}
\begin{center}
\epsfxsize=5in\epsffile{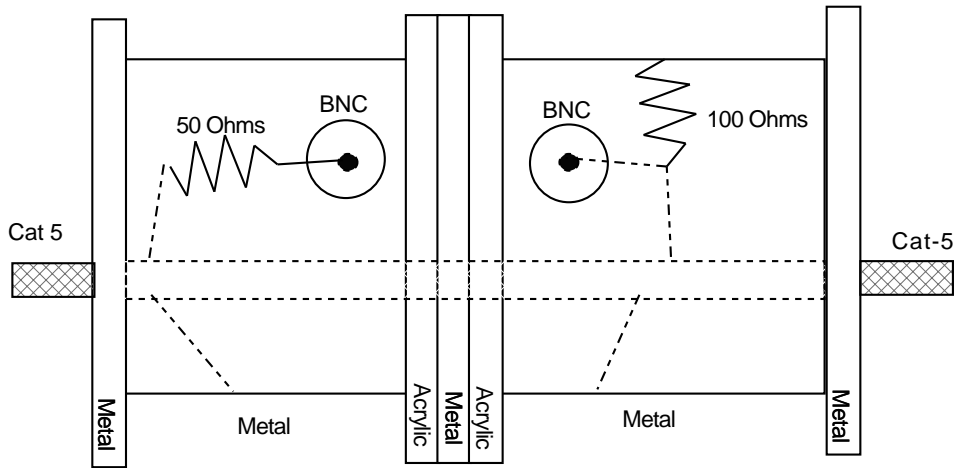} 
\caption{Cross Cable Impedance Matched onto RG-58/U Coax, and diagram of
the shielded box}
\label{fig2}
\end{center}
\end{figure}

As such, the hardware requirements should be quite easy to meet for 
even the most restricted budgets. Again, when the 
experiment is finished or otherwise not in use the cables find
use in connectivity in the rest of the lab/school. 

\begin{figure}
\begin{center}
\epsfxsize=5in\epsffile{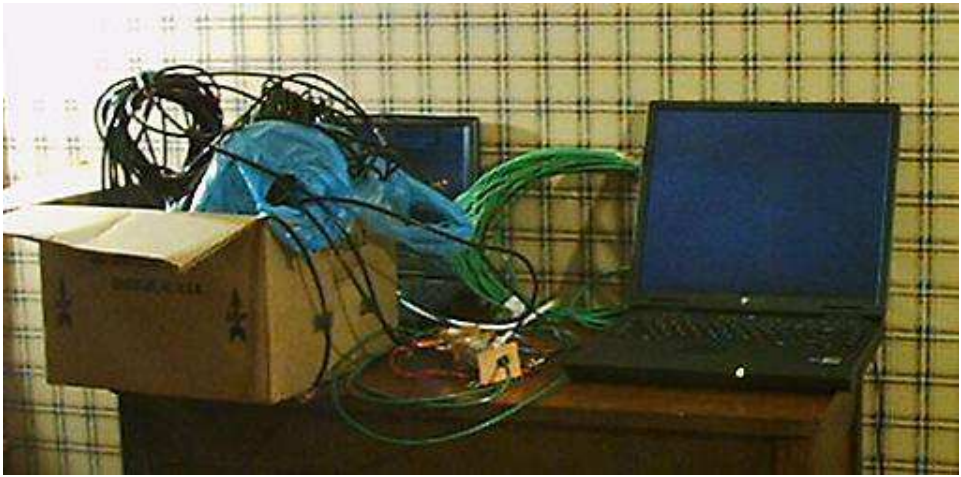} 
\caption{Picture of the Laboratory Setup during coax runs.}
\label{fig3}
\end{center}
\end{figure}

{\bf Software:} We took data while running Linux 
on both computers. Although it should be possible to do this 
experiment  
with the new release of ping for Windows,  because the 
authors were unfamiliar with Windows, Linux was chosen. 

The Linux operating systems (OS) used (both Red Hat and Slackware) 
are free and widely distributed. These releases come with a
system utility called {\it ping} which basically broadcasts
a small (usually less than 100 bytes) IP packet through a
specific internet adapter to another machine (specified by its 
IP address) on the network. That machine basically acts as a mirror. 
The packet header identifies the packet as a 
{\it ping} command and a process running on the
'mirror' identifies it and rebroadcasts it (back) to the sender. The sender
than checks the reflected packet for errors and reports the 
total roundtrip time. 
 {\it Ping} has evolved much over the years; in the current Linux releases
it reports the roundtrip time to the nearest microsecond whereas
older versions only return the 
roundtrip time to the nearest millisecond. 
In this experiment we used the current version of {\it ping}. 

Here is an example of the output of a basic {\it ping} command, in this case a
computer {\it ping}-ing itself (the very first line (bold) 
was actually the command entered, the rest is output). 
\bigskip

$>$ {\bf ping 168.192.0.10}

PING 168.192.0.10 (168.192.0.10) from 168.192.0.10 : 56(84) bytes of data.

64 bytes from 168.192.0.10: icmp\_seq=0 ttl=255 time=576 usec

64 bytes from 168.192.0.10: icmp\_seq=1 ttl=255 time=82 usec

64 bytes from 168.192.0.10: icmp\_seq=2 ttl=255 time=78 usec

64 bytes from 168.192.0.10: icmp\_seq=3 ttl=255 time=78 usec

64 bytes from 168.192.0.10: icmp\_seq=4 ttl=255 time=76 usec

--- 168.192.0.10 ping statistics ---

5 packets transmitted, 5 packets received, 0\% packet loss

round-trip min/avg/max/mdev = 0.076/0.178/0.576/0.199 ms

\bigskip
Each of these five packets have been sent about one second apart. 
The transcript always includes a summary (at the end) indicating
how many packets completed the reflection/reception successfully and, 
importantly,  gives some gross statistics for the pings including the 
average round trip time to the nearest microsecond. 

We found four switches, the {\it -f} the {\it -i} the {\it -q} and the
{\it -c}, of 
{\it ping} quite useful for this experiment. 
The -f or 'flood' switch instructs the {\it ping} daemon to keep 
sending out a new packet as soon as an earlier packet is received back. 
Because this 
can put some strain on a real network this 
command can only be executed by a user with {\it root} 
or 'superuser' privilege, although for our two-computer network this
is not a concern. 
The command {\it ping -f $<$IPaddress$>$ } is very useful in that 
it can quickly tell you whether there is enough fidelity on a 
particular cable to warrant running the experiment. If {\it ping -f} 
returns with many dropped (non-returned, or lost) packets OR if 
multiple applications of it reveal that there is not a steady (up to 
$\pm$ 1 microsecond) average return time, then the cable may not be 
suitable for this experiment or there are other sources of drift which will 
complicate the data collection/analysis of the experiment. 

The command {\it ping -i .01 $<$IPaddress$>$} 
issues a new packet every .01 seconds. This is used in the data collection 
since it it allows a student to accumulate on the order of 
one hundred thousand of pings in about 15 minutes. Large datasets 
are a necessity for subthreshold signal detection, and going 
to ten or 15 minute data collection times allows sufficient averaging 
of the noise so that signals ({\it i.e.} delays) 
on the order of 1/20$^{th}$  of the threshold
({\it i.e.} about 50 nS) can be reproducibly resolved, yet in short enough 
time that the data acquisition can be completed in a lab period. 

{\it Ping} reports time in microseconds, but 
it is not clear {\it a priori} how well 
its time measurement is 
calibrated. 
In order to calibrate the roundtrip times reported by {\it ping} 
we found the {\it ping -q} and {\it ping -c} commands quite useful. 
The switch {\it -q} simply forces ping to run quietly until the end when 
it writes its summary data to the 'standard I/O' (the users screen). This
means that while the system is running {\it ping} it has only the 
overhead associated with acknowledging the packet reception and issuing 
the next packet transmission. The rest of the time the system is waiting 
for the packet to return. Finally, the {\it -c \#N} tells pings to run 
for exactly {\it \#N} send-and-receive cycles. 

It is easy to 
calibrate the {\it ping} times by combining 
the {\it ping} switches described above with the Linux {\it time}. The 
{\it time}  
command returns the system time, user time and overall time it takes for 
a command to complete. For example, for the particular experimental 
configuration that we used (whose full specification is below in the 
``Case Study'' section) we issued the command (as {\it root} on the 
fast machine as described in the Experimental Design section below) 
``{\it time ping -rfqc 300000 $<$IPaddress$>$}'' where {\it $<$IPaddress$>$} 
was the 
IP number of the other machine. This runs the 'flood'-style ping for 
exactly 300000 times between the machines, 
returns summary data both from {\it ping} and from {\it time}. 
There is a noticeable (at the few percent level) 
difference between the reported ping times
and the actual run time of the command, and in the case study below we use
the difference to calibrate {\it ping}~'s microseconds.

Finally, in order to insure prompt packet handling, it is necessary
to configure
our machines so that they are on a common IP subnet. 
For example, if the machines are
setup with a subnet mask of (the traditional) 255.255.255.0, then both machines
must have the same first three numbers in their IP address, for example, 
168.192.0.13 and 168.192.0.14. 

\bigskip
{\bf Experimental Design:} The basic idea of the experiment is to 
record round-trip {\it ping} times on two machines connected by a cable 
and to analyze how the average round-trip time changes when the 
cable's length is changed. 

In the actual experiment the computers and cables 
were not moved. 
Their batteries were
fully charged and they were continuously plugged into wall
power source for the entire duration of the data collection. 
After allowing them 
to warm up for about an hour while pinging {\it ping -f} 
and {\it ping -i .01} to make sure the card and line were working properly, 
the mirror machine was reset so that there were no extraneous
processes running on it (no users logged in or any batch activity other 
than what the basic system requires). Similarly, the transceiver 
machine was dropped into a text shell as 
{\it root}\footnote{It was discovered that with 
``X'' or other desktop environment running on either machine the 
overhead associated with such an interface introduced additional 
timing noise which complicated the simple analysis we present here. 
One of the universal features of
stochastic resonance is that the signal first increases with 
noise amplitude, and then decreases as the noise amplitude increases. 
No noise doesn't work, a little does well and too much swamps out 
the signal altogether!}. 
This leads to a stable enough environment for the machines and their 
network that the data analysis is relatively straightforward (see below). 

The data collection then commenced with a {``{\it ping -i} .01
$<$IPaddress$>$~ $>$ tempfile1.txt'' where ``$>$'' 
(the so called 'pipe' symbol) 
in unix/Linux means that the output is redirected to the disk
file ``tempfile1.txt''. This was allowed to run for 11 minutes
or so and then stopped, another cable length inserted and the command 
rerun, now with the data piped to a (differently) named file
({\it i.e.} ``tempfile2.txt''), and so 
on, until  a series of runs were completed with various cable lengths.
Ten runs with three or four 
different cable lengths can quite easily be completed in a two-hour 
laboratory period. In principle, staggering the order of cable lengths 
tested allows one some control of any systematic drift in the 
equipment over time. In practice that allowed us to verify
that drift over the course of two hours 
was smaller than the resolution (~50 nS) 
and so could be ignored, further simplifying the data analysis. 

Each raw data file ('tempfile.txt' in the description above) is a few 
megabytes. It is processed so that it ultimately 
consists of just two columns, the attempt number and the measured 
round-trip time for that ping attempt. Simple cuts are made (see details 
in the case study below) and the average
and standard deviation of the remaining roundtrip times is computed and 
recorded. This analysis is straightforward enough to 
be done in a spreadsheet. 
Finally, the roundtrip distance (or cable length for the 
case of the coaxial cable) is plotted against average roundtrip time and 
(least-squares) fit with a line, the slope of which is the measured
speed of signal propagation.

A more subtle point has to do with the properties that the timing noise
must possess in order for it to faithfully assist the 
subthreshold signal detection. The relevant {\it ping} return times
typically form a cluster about a few microseconds wide on a typical 
return time of hundreds of microseconds (see Figure 4). 
Thus, as long as the distribution 
of timing noise is relatively flat near the hundred microsecond band, the 
noise will faithfully assist in subthreshold signal detection. That this 
is indeed the case is evidenced by the reproducibility and data quality
of the runs, allowing the
reproducible 
average round-trip time resolution of about 50 nS 
(or 1/20th of {\it ping}'s reported resolution).

\bigskip
\noindent{\bf III. Case Study Example and Data Analysis Summary}

Figure 2 shows a typical graph of round-trip times for the setup 
described above (in this case we use a 3.53m cat-5 crossover 
cable between a Gateway
Solo 9500 (950 MHz with 100BT ethernet built in) running Red Hat Linux 7.1 
and a Gateway Solo 2300 (with a 10BT 
3COM589 PCMCIA slot ethernet card) running Slackware Linux 2.0.35).
As such, note that the cable actually is running at 10BT speeds. 
We did however
find similar data and data quality on other combinations of machines and 
speculate that the particular hardware doesn't matter, as 
long as the machines are not identical (which might complicate the stochastic
resonance).  
\begin{figure}
\begin{center}
\epsfxsize=5in\epsffile{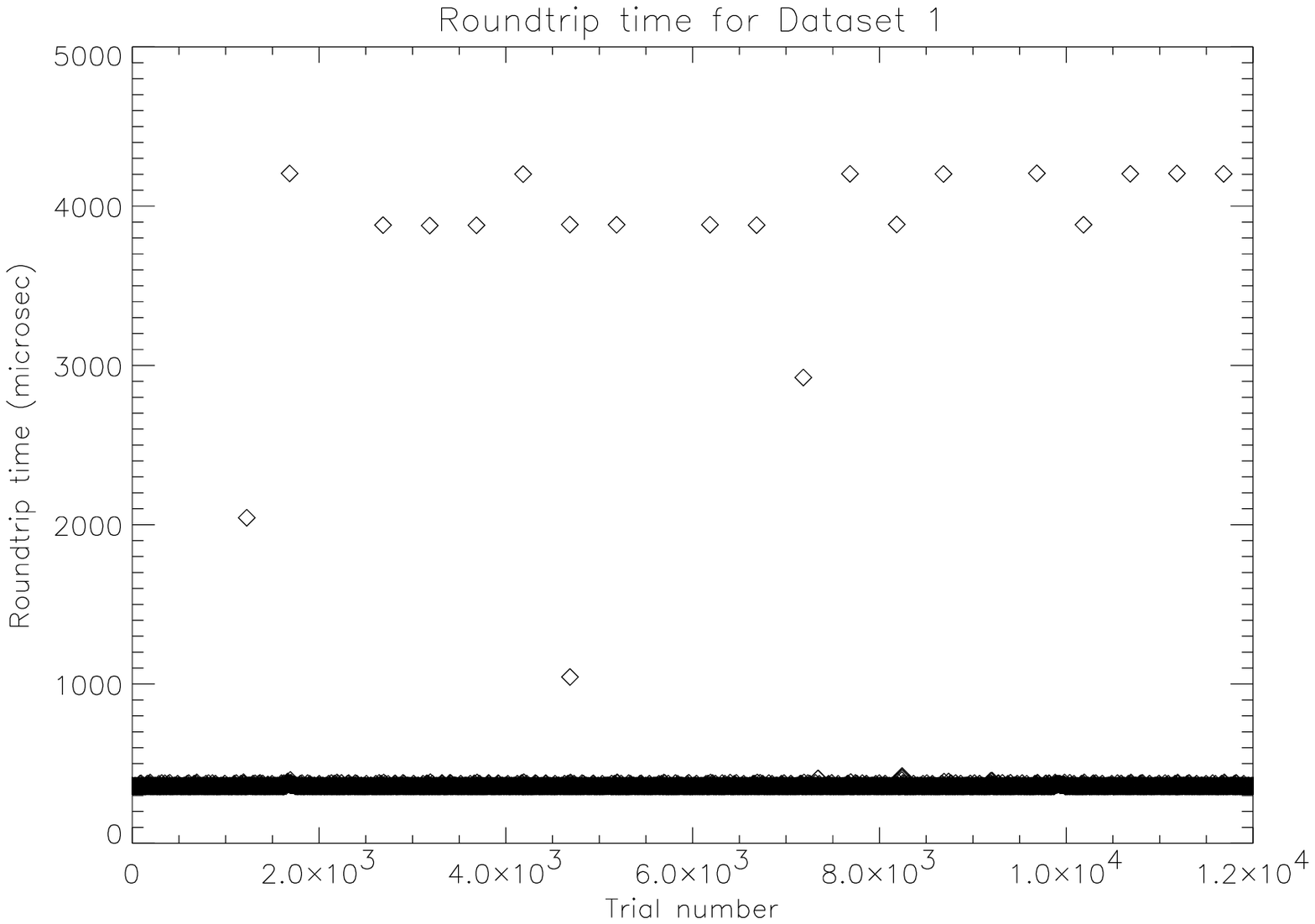} 
\label{fig4}
\end{center}
\end{figure}
\begin{figure}
\begin{center}
\epsfxsize=5in\epsffile{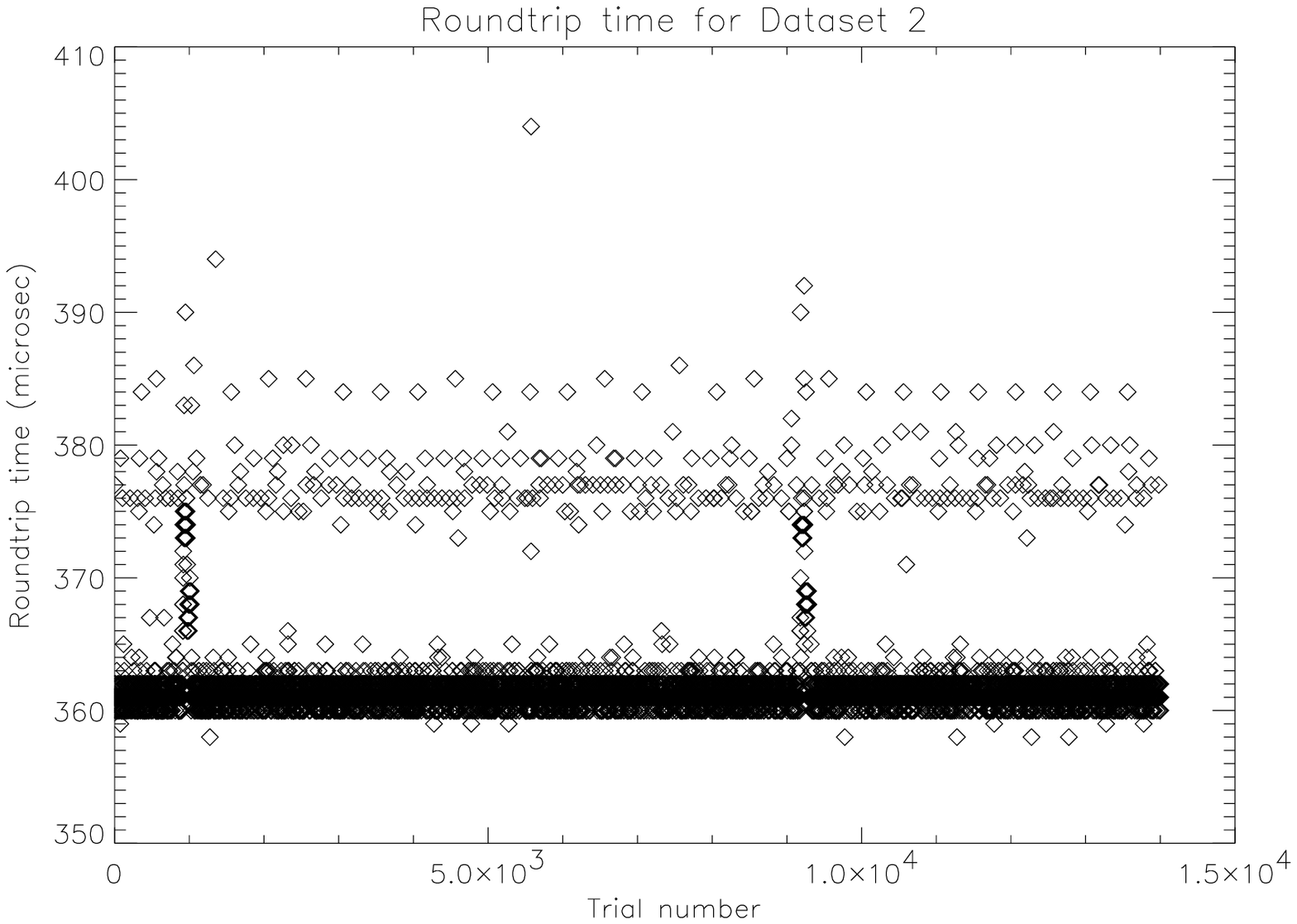} 
\caption{Graph of 12,000 round-trip times as a function of 
attempt number, and detail}
\label{fig5}
\end{center}
\end{figure}
Note that most of these times cluster around 360 microseconds with a
substantial set of outlyers, with some packets taking several milliseconds
to return. The detail of the region around 360 microseconds indicates
a spread in the main part of the data and characteristic timescales 
associated with longer recorded times. Additionally, there are 
periodic structures seen in these data; both the longer recorded times 
and the periodic structures we regard as extraneous; being the 
result of periodic and aperiodic disturbances associated with scheduled 
low-level tasks that are running in the OS. 
A histogram of the round-trip times of  Figure 2 is shown 
below (Figure 3). 
\begin{figure}
\begin{center}
\epsfxsize=5in\epsffile{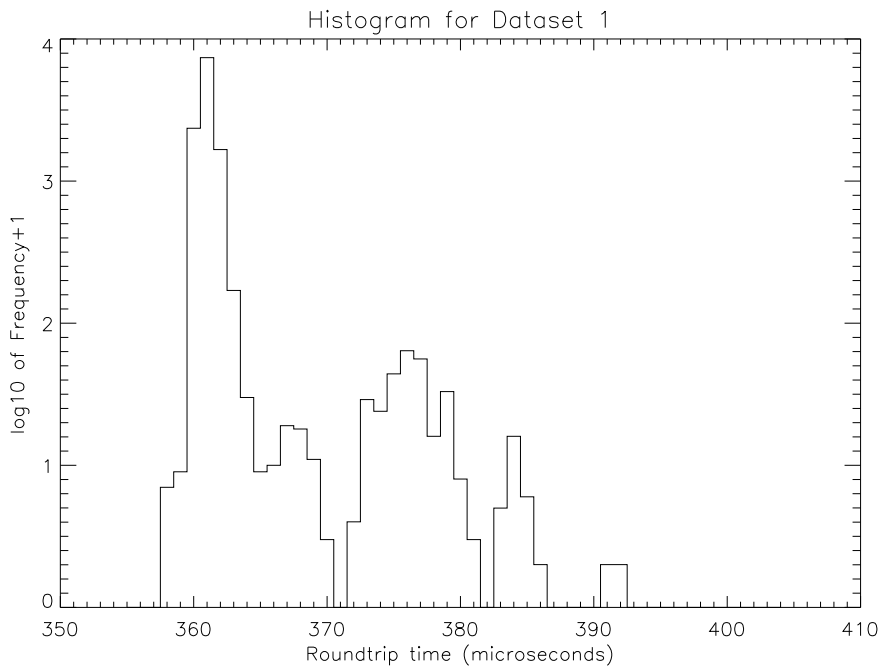} 
\caption{Same 12,000 roundtrip times in a log(frequency) histogram}
\label{fig6}
\end{center}
\end{figure}
These datafiles are then stripped 
of outlyers. 
We make a cut, for the example given above, keeping only those {\it ping} 
round-trip times of 365 microseconds or less and analyse all of the 
data sets in these runs with this same cut. 
(We found that the choice of the cut did not substantially affect 
our measured propagation speeds, 
the sensitivity to reasonable but different choices of cuts
being at about the 0.5 percent level. 
Also, although we did not have a lower cut on the data, each million 
or so {\it ping}s we occasional found a return time logged as 0
microseconds. These very rare spurious lines were 
also removed from the datasets. 
We used {\it Interactive Data Language} (IDL$^{tm}$) for the data analysis, 
fitting and plotting. 
The data analysis steps are elementary and may also be easily automated
for an introductory class setting, or could be done on a spreadsheet or
easily in {\it Maple}$^{tm}$ or {\it Matlab}$^{tm}$. 
For the data set above we 
find that the data so filtered has a reported average round-trip time of 
360.12 microseconds and a standard deviation that is about .64
microseconds, an error figure much bigger than the resolution we are claiming. 
It is due primarily to the width of the bins that the data is in, and 
is not a good measure of how well one can find the center of 
(an assumed nearly) gaussian distribution from these coarse data.

To calibrate the time reported by {\it ping} we issue a 
{\it {time ping -rfqc 300000 $<$IPaddress$>$ }} from the prompt and
find an average roundtrip reported packet time of 369 $\mu$S
and so the reported time spent waiting for the packets to return is
300000 $x$ 369 x 10$^-6$ = 110.7 seconds. The {\it time} command
returns a total 
time of 111.3 seconds, of which 4.02 seconds are spent in system
calls for both the maintenance of the OS (which wasn't doing any 
extraneous jobs) and the low-level {\it ping} system calls. This means 
basically 107.3 seconds were spent waiting for the {\it ping}~s to return. 
Thus, the actual reported {\it ping} times were longer than the actual times
by about 3\%. We include this systematic error in our determination 
of the speed of propagation in the data analysis that follows. 

Having done the same cut and tally with datsets for all 
five cable lengths (adding a length of .75, 3.07, 15.03, 37.5 and 57.15 
meters to the gender change cross cable)
we plot twice the category-5 cable length versus the average
round-trip times in Figure 5
\begin{figure}
\begin{center}
\epsfxsize=5in\epsffile{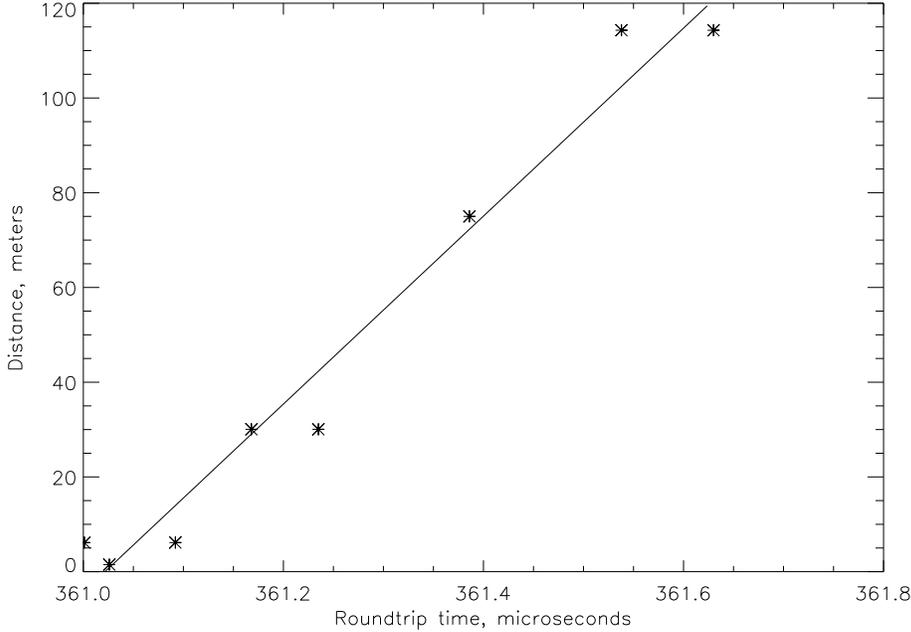} 
\caption{Round Trip distance (meters) versus round trip time (microseconds) 
in Cat-5 cable.}
\label{fig7}
\end{center}
\end{figure}
The least squares fit to these data yields a 
slope of 2.04$\pm$.14 x 10$^8 ~{{m}\over{s}}$, indicating that the 
speed of propagation in a cat-5 cable is some 2/3 the speed of light 
in the vacuum. 

Finally, in order to use this technique to estimate the speed of light in 
the vacuum we need some way of converting our cable measurements. This 
can easily be done with elementary electromagnetics theory
for simple 
cable geometries. As a first approximation, the twisted pair in 
category 5 cable can be modeled 
as an (untwisted) pair of parallel wires each of radius 
$r$ completely surrounded by a 
dielectric material (the insulation, dielectric constant $\epsilon$) 
and with their centers held a distance $d$ apart. 
The capacitance per unit length of this system can be found 
by a straightforward application of the method of images, and the 
inductance can be found by integrating the magnetic flux between two 
current-carrying wires. 
The capacitance and inductance per unit length for such a transmission 
line is thus a standard computation yielding, 
\BE
c = {{\pi \epsilon \epsilon_0}
\over{\log\bigl( {{d}\over{2r}} + \sqrt{ {{d^2}\over{4r^2}}-1}
\bigr)}}
\qquad \qquad 
l = {{\mu_0}\over{\pi}} \log\bigl({{d-r}\over{r}}\bigr) +  
{{\mu_0}\over{4\pi}} \rho(\omega)
\label{cl_wirepair} 
\EE
with $\rho(0) = 1$ and $\rho(\infty) = 0$ is a function associated with 
the skin effect in the conductor. At the frequencies of interest here
$\rho \sim 1$. 
The telegrapher's equation indicate that the 
speed of propagation ($v=1/\sqrt{cl}$) 
for electromagnetic waves along the 
transmission line and line impedance ($z=\sqrt{ {{l}\over{c}}}$)
for this cable are,
\BE
v = c_0  \sqrt { {{\log\bigl( {{d}\over{2r}} + \sqrt{ {{d^2}\over{4r^2}}-1}
\bigr)}\over{\epsilon ~\log\bigl({{d-r}\over{r}}e^{1/4}\bigr)}}}
\qquad \qquad
z = {{z_0}\over{\pi}} \sqrt { {{1}\over{\epsilon}} 
\log\bigl( {{d}\over{2r}} + \sqrt{ {{d^2}\over{4r^2}}-1}\bigr)
~\log\bigl({{d-r}\over{r}}e^{.25}\bigr)}
\label{speedwirepair}
\EE
Where $c_0$ and $z_0=377 ~\Omega$ are the speed of light in the vacuum and the 
impedance (in Ohms) of free space respectively. 
These expressions can be compared to other, more approximate
expressions in the literature such as 
$c=0.276 ~{\rm x}~ 10^{-10} \epsilon/\log(d/r)$, 
$l=0.400 ~{\rm x}~ 10^{-6} \log(d/r)$ and 
$z = 120.\log(d/r)/\sqrt{\epsilon}$ from Ref.\cite{radlab}. 

For cat-5 cable, $r$ = .00026m and $d$ = .00086m. 
The DC capacitance per unit length of a pair in cat-5 cable is measured
(by a DVM) to be about 45pF/meter, 
so this indicates that the dielectric constant
at low frequency is $\epsilon = 1.76 $ where we have used  
Eq.~(\ref{cl_wirepair}). Thus the line impedance 
is computed by Eq.~(\ref{speedwirepair}) to be
$z = 98 \Omega$,  
in rough agreement with the specifications for cat-5. 
Assuming this is the same dielectric 
constant at radio (ethernet) frequencies, we use Eq.~(\ref{speedwirepair})
and our data thus imply a speed of light in vacuum of 
$c_0= 2.70 \pm .18 ~{\rm x}~ 10^8 {{m}\over{s}}$. 

The main error in this determination of the speed of light in a vacuum
using cat-5 cable data seems to be systematic effects associated 
with deficiencies in our model. We now estimate these effects. 
The twisting reduces the line impedance (chiefly by 
increasing the capacitance per unit length through the 'extra' metal that
the full twists effectively act as metal loops around the other conductor) 
and increases the actual length of the line as compared with the above model 
of straight parallel wires. A twist length of .7 cm in these cables
(see Appendix A) indicates that to straighten out the wires in a pair 
we must 
twist the wires around each other 140 
times each meter.
Secondly, the conductors are not submerged in 
dielectric, but the dielectric is actually quite thin and so use 
of Eq.~(\ref{cl_wirepair}) is suspect. The net effect of these model 
deficiencies is to reduce the speed of signal transmission as 
compared with two parallel conductors. Even between two pairs 
(with different twist length) 
in the 
same cat-5 cable there 
can be as much as a 10 \% effect due to the twist length of the pair, with 
more twists per unit length associated with a smaller overall speed. 

We can very approximately account for the systematic 
effect due to extra length caused by twisting of the lines.
This is easy for students to correct for, either by actually comparing the 
lengths of a pair before and after straightening out or 
geometrically from the measured 
data above; the actual length $L$ of conductor inside a twisted pair 
of length $L_0$ is simply $L = 
L_0 \sqrt{1 + \bigl({{\pi d}\over{l_{tw}}}\bigr)^2}$ where
$d$ is as before and $l_{tw}$ is the twist length of the pair. For a typical 
cat-5 cable the computed 
actual length $L$ of the pair is some 
(for .7cm twist length) 7.2 percent longer than the cable length. Amending 
our untwisted estimate above by this 'geometrical' correction, our 
measurement indicates that $c_0 = 2.90 \pm .19 ~{\rm x}~ 10^8 {{m}\over{s}}$, 
a 7 percent sigma, and well within one sigma from the defined value for    
the speed of light in the vacuum. 

Although this is a reasonable outcome, 
in an attempt to simplify this laboratory by avoiding 
some of these dicey geometrical and 
electrical intricacies of cat-5 cables, we repeated the above 
laboratory in a geometrically simpler cable, RG-58/U, 
otherwise known as standard 50 $\Omega$ coaxial cable. 
We already described in the hardware section efforts we made to 
impedance match and shield the cable junctions. 
The radius of 
the inner conductor of RG-58/U is measured to be 
$r$= .00083m and the inner radius of the sheath is 
$R$=.00296m. For a interconductor dielectric $\epsilon$ we can again 
resort to elementary electromagnetics theory to ascertain that the 
capacitance and inductance per unit length of a coaxial cable is
\BE
c = {{2 \pi \epsilon \epsilon_0}\over{\log(R/r)}} \qquad \qquad 
l = {{\mu_0}\over{2\pi}} \log(R/r) + {{\mu_0}\over{4\pi}} \rho(\omega)
\label{cl_coax}
\EE
(compare with Ref.\cite{jackson})
Which again may be compared again with the perhaps more approximate results
$c=0.555 ~{\rm x}~ 10^{-10} \epsilon/\log(R/r)$ and 
$l=0.200 ~{\rm x}~ 10^{-6} \log(R/r)$ from Ref.\cite{radlab}. 
These analytical expressions lead to
a speed of propagation and a line impedance of 
\BE
v = c_0/\sqrt{\epsilon}+\ldots \qquad \quad 
z = {{z_0}\over{2\pi \sqrt{\epsilon}}} \log(R/r)
\label{speedcoax}
\EE
where $c_0$ here is the speed of light in a vacuum. The $\ldots$ 
is associated with the second term in Eq.~(\ref{cl_coax}) for $l$.  
The measured DC capacitance per unit length of our cable to 
is  97 pF/m (see also Ref.\cite{leo}, pg. 256). Eq.~(\ref{cl_coax}) thus
indicates that the DC dielectric constant is 2.22, and, as a check, 
these data imply that the line impedance is $52 \Omega$. 

Four coax lengths, .333m, 11.61m, 25.0m, and 74.5m were used. 
The cuts were performed, averages taken 
and Figure 5 is a graph of the coax cable length (since only one of 
the twisted pairs lines is of variable length) versus average
packet round-trip time. 
\begin{figure}
\begin{center}
\epsfxsize=5in\epsffile{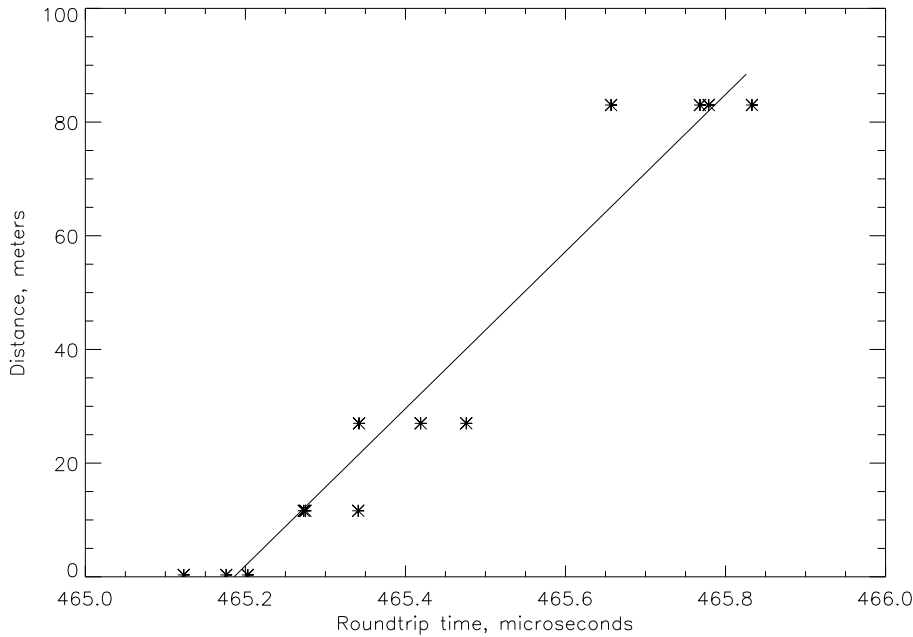} 
\caption{Coaxial cable length (meters) versus trip time (microseconds).}
\label{fig8}
\end{center}
\end{figure}
Thus the measure speed of propagation of an electromagnetic 
wave in RG-58 coax is 1.27 +/- .09 x $10^8 ~{{m}\over{s}}$, over
seven meters per nanosecond. This is substantially  different than the 
$.57c_0$ expected by the above formulas. 

It may be that reflections and attenuation 
in the line play an important role in understanding this
discrepancy. Although we shielded and attempted to impedance match 
the cables, there may still have been enough reflections at the 
connectors to cause phase shifts that may have delayed the signal 
acquisition by the ethernet cards. Also, the matching circuit that 
we used causes about 75\% of the signal power to be lost at the junction 
box. The reduced signal also suffers further diminution in transit. 
If an attenuated signal takes longer for the ethernet cards 
to detect this will be a systematic effect that would reduce the 
measured propagation velocity. Such an effect may not have been 
seen in the cat-5 sections because the ones we used were so 
short that at 10BT speeds there should have been less than 10dB 
of loss in even the longest cable.

\bigskip
\noindent{\bf IV. Conclusion and Critique} 

Using {\it ping} to measure the speed of propagation of 
electrical signals in a cable and thus to determine the speed of 
light works reasonably well (at the 7\% level) in cat-5 cable. 

For simplicity of execution and analysis, low cost, dual use
and accuracy
this laboratory is hard to top! Since this lab does not make
use of any of the wave properties of light it will be simpler for 
students to do earlier in their second semester of physics. 
It does make use of basic internet-style resources that students
may find useful later, and may stimulate students to think 
more about the usefulness of stochastic resonance phenomena.

Of course, the disadvantage of this method is that the student 
has to marshal some theory both conceptually (that electrical signals 
in a wire are like light) and analytically to relate their 
measurements to the measurement of the speed of light in the vacuum.  

Our development of this laboratory indicates points for further 
study. We have not yet been able to definitively determine why
the method works so poorly in coax, though we have some indications 
that the junction box we used may be part of the problem. An 
interesting observation in that regard is that we actually tested
ganging coax cable lengths with the use of a barrel connector and 
found that the presence of the barrel significantly 
distorted the measured return times. Apparently barrels are not 
so well matched onto 50 Ohm coax. In any case, if one wants to 
build the lab based on the geometrically simpler RG-58/U 
cable, it is possible to buy network cards that are made for 
this cable and come with BNC female connectors installed. 

Also, this all raises the obvious question as to why we didn't 
redo the experiment with a wireless network card and hub. These 
can be bought at a modest cost and we did try to do this laboratory 
with wireless ethernet. However, due to peculiarities with small 
packet handling in the wireless protocol we used (IEEE 802.11b) 
the return times were far too long and noisy for stochastic resonance
to work. We have not yet looked into  doing this with IR links 
(which may be too short) or fiber links, but that is also a logical 
next step.

\bigskip
\noindent {\bf Acknowledgments} We are thankful to Mark Welton and Ray Hoff
for discussions on ethernet and cabling and to Jeff Carroll and the 
XEL team at Y.S.U.'s Center for Photon Induced Processes for the RG-58/U 
cable lengths used in the experiment. 

This work was supported in part by a grant from the DOE EPSCoR grant
DE-FG01-00ER45832, Research Corporation Cottrell Science Award \#CC5285, 
a Cluster Ohio Grant from the Ohio Supercomputer Center, NASA grant 
NAG9-1166 and a Research Professorship 2001-2002 award from YSU. 
This work was (partially) supported by the National Science Foundation 
through a grant for the Institute for Theoretical Atomic and Molecular 
Physics at Harvard University and the Smithsonian Astrophysical Observatory. 


\newpage

\ \

\newpage
\ \ 
\vskip .1in 
\centerline {\bf Appendix A: More about Network Cables}
\vskip .2in

Invariably a laboratory like this opens the classroom up 
to all sorts of discussions about what is 'really going on' and 
about particulars of the physics of wave propagation in the 
physical fabric of the internet. In this appendix we have 
collected information about cables and computer networks that 
an instructor may find useful to consult for answers to some 
of the questions that students may ask 
as a result of doing this
laboratory. 

In the first part of the experiment cat-5 cable was used. 
It consists of 8 lines in 4 unshielded twisted pairs (UTP) 
and is terminated by a male RJ45 connector, which looks like a 
wide phone jack. 
The only pairs of direct relevance for our experiment are the 
green (green and white-green lines) and orange 
(orange and white-orange lines) 
pair. One of the pair (pins 1 and 2) 
is for transmission and one 
is for reception (pins 3 and 6). 
For the purposes of this experiment, all the cables 
could have been of cross-cable type, that is, the green pair and the 
orange pair are interchanged between the two RJ45 connectors (thus 
connecting a transmission on one side of a cable to the  receiver 
on the other end; See for example Figure 1 in the text). With such 
a cable arrangement it is possible to connect and run IP from one 
ethernet NIC'd computer to another, but only two computers may be 
so connected. 

In order to join more than two computers to one another it is 
imperative to use either a hub or a switch. Besides cost (hubs 
can be bought for less than \$20, a cheap switch can be had for \$50) 
the main 
technical difference between hub and switches is intelligence; 
The hub just 
acts like a dumb repeater and rebroadcasts the packet received across 
all lines (thus occupying all lines for each packet on the network) 
whereas 
a switch can route a given packet to a particular plug/cable associated
with the destination machine. The switch does so by reading and decoding 
parts of the packet headers (sort of a preamble to the actual data content 
of the packet) which identify the IP numbers of the  machines accessible on 
each wire and contains also the IP number of the destination machine
and many can do so in at least a partially ``non-blocking'' 
fashion where mutually exclusive pairs of machines can simultaneously 
communicate.
Because of the reduced network load of switching, hubs are not typically 
used except in local, small clusters of machines. 

Typically, IP packets are smaller than about 20 Kbytes
(this is in a protocol called
IP version 4); the {\it ping} packets used in this experiment are 
quite small, being less than about 100 bytes. When you send a message 
bigger than 20 Kbytes with IP it is first chopped up into packets of 
about 20 Kbytes each, and each packet is labeled 
as if it were a multi package 
mail order, that is , 'packet one of 85' for the first one, 'packet 2 of 85'
for the second one and so on. The packets are sent independently, and thus may
take 
different routes to the destination machine. Thus they may 
arrive in some jumbled 
temporal order but are reassembled into a whole at the destination 
computer after the last packet arrives. 

The cat-5 cables are themselves actually highly engineered products. The 
pairs of UTP in the cable are each twisted at different incommensurate 
twist lengths, typically varying between .6 and .9 cm. Additionally, the 
pairs themselves are twisted about each other. The rapid, incommensurate
twisting in each pair and the twisting as pairs all reduce the amount of 
cross talk (signal from one pair bleeding over into another pair in the 
cable) and RFI (radio frequency interference from sources outside the 
cable) experienced by waves in the cable pairs. A typical cat-5 cable can 
transmit a 100 MHz signal a distance of 100 meters with a loss of signal 
of roughly 20 dB, increasing with frequency to about 40dB at 300 Mhz. 
These frequencies are not atypical of 100baseT connections, though 
10baseT connections work at substantially lower frequencies. Actually, it is 
not the attenuation (signal loss) that prevents one from running IP 
over cables longer than 100m, but the protocol's internal packet collision
handling specification that does. There are many good descriptions of this 
on the web, including the {\it wild packets} site 
(see Ref.~[\cite{webresources}]). 

Physically, the smaller the twist length on a pair 
the larger the capacitance per 
unit length. This reduces the speed of propagation. Since the pairs all 
have incommensurate (but fixed) twist lengths, the signals sent at the 
same time reach the other end at different times. This undesirable 
effect, called
{\it skew}, can be as large as 10\% of the total cable transit time. 
One technique to make low-skew cables involves
making wire pairs with different 
insulation dielectric constants so that the pair with the shortest 
twist length has the smallest insulation dielectric constant. 

Older category-3 cable is also comprised of UTP but 
differs from cat-5 primarily by the twist length.
Typical twist lengths in cat-3 are from 8 to 10 centimeters.  
As a result, cat-3 cable can only support transmission frequencies below about
16MHz, and in our tests, because these cables have much more cross-talk and 
are susceptible to RFI, we were unable to get reproducible average roundtrip 
times from cat-3 cable with the procedure described in this paper. 
The published data for cat-3 does indicate that, 
as expected from the above comments, the speed of signal propagation is 
significantly higher ($~>.7c$) than cat-5 ($~.66c$) though some 
cat-5(e) cables do also have propagation speeds nearly as high as cat-3. 
For students, discussing the 
difference between cat-5 and cat-3 cable will dramatically 
highlight the important physical difference 
between speed of propagation and bandwidth. 

As a further 
illustration of this, optical fibers
(multimode) typically also have slower propagation speeds yet (about .6c) 
but have 
over a million times the bandwidth of cat-5. In many installations 
the cat-5 lines from your client
go to a cabinet where they are connected by a switch onto a single 
fiber 'trunk' line that brings in IP service to the entire installation. 

The impedance per unit length of both cat-3 and cat-5 pairs is about 
100 Ohms. RG-58/U coax has an impedance of 50 Ohms. The impedance is 
the ratio of the voltage to current of a traveling wave on the 
transmission line. At a 
junction of two cables with unequal impedances a pulse incident on the 
junction will be partially reflected; the amount of reflection will be 
precisely that necessary to ensure that the resultant 
traveling waves in each cable 
have the requisite voltage/current ratio. This is the same as seeing your
reflection in a window. Such reflections are undesirable for this experiment 
and so were greatly reduced (though perhaps not eliminated) 
through the impedance matching resistors 
of Figure 1. 

The speed of propagation also depends in principle on whether the 
cable is straight or wound. All our cables (except the 
very shortest ones) were wound on a radius of about .25 meter. The 
effect is noticeable in some cable types, notably in RG-59, 
the 75 Ohm coax. A very readable discussion of this effect for various
cable types can be found at the JTE site http://www.nteinc.com/jovial/. 
These data indicate that the effect on the propagation speed 
as a result of coiling the cables is completely negligible for 
cat-5 and RG-58/U cable.




\end{document}